\documentclass[english,english,noshowpacs, amssymb,twocolumn,aps,nofootinbib]{revtex4}
\usepackage[T1]{fontenc}
\usepackage[latin9]{inputenc}
\usepackage{color}
\usepackage{mathtools}
\usepackage{textcomp}
\usepackage{amsmath}
\usepackage{esint}
\usepackage{url}
\usepackage{natbib}
\usepackage{graphicx}
\usepackage[export]{adjustbox}

\makeatletter

\newcommand{\lyxmathsym}[1]{\ifmmode\begingroup\def\b@ld{bold}
  \text{\ifx\math@version\b@ld\bfseries\fi#1}\endgroup\else#1\fi}

\@ifundefined{textcolor}{}
{%
 \definecolor{BLACK}{gray}{0}
 \definecolor{WHITE}{gray}{1}
 \definecolor{RED}{rgb}{1,0,0}
 \definecolor{GREEN}{rgb}{0,1,0}
 \definecolor{BLUE}{rgb}{0,0,1}
 \definecolor{CYAN}{cmyk}{1,0,0,0}
 \definecolor{MAGENTA}{cmyk}{0,1,0,0}
 \definecolor{YELLOW}{cmyk}{0,0,1,0}
}
\makeatother

\usepackage{babel}
\begin{document}

\title{Comparison of coherence area measurement techniques for bright entangled twin beams}

\author{Ashok Kumar{\footnote {ashok@ou.edu}}, Hayden Nunley, and  Alberto M. Marino{\footnote {marino@ou.edu}}}
\affiliation{Homer L. Dodge Department of Physics and Astronomy, The University of Oklahoma, Norman, Oklahoma 73019, USA}

\begin{abstract}
Quantum states of light with multiple spatial modes are fundamental for quantum imaging and parallel quantum information processing. Thus, their characterization, which can be achieved through measurements of the coherence area, is an important area of research. We present a comparative study between two different measurement techniques for the coherence area of bright entangled twin beams of light generated with a four-wave mixing process in a hot rubidium vapor cell. The first one provides a direct characterization of the size of the coherence area and is based on correlation measurements between spatial intensity fluctuations of the twin beams with an electron-multiplying charge-coupled-device camera. The second one provides an indirect measure  and is based on a noise analysis of different spatial regions of the twin beams in the time domain with a single photodiode.  We show that the indirect technique, which can be implemented with a significantly less complicated measurement scheme, gives an estimate of the size of the coherence area consistent with the direct measurement technique performed in the spatial domain.
\end{abstract}

\maketitle

\section {Introduction}

Spatial quantum correlations between entangled photons or bright twin beams have gained significant attention due to their application in testing fundamental quantum physics~\cite{RMP}, quantum imaging~\cite{Kolobov,Brida}, quantum metrology~\cite{Giovannetti} and quantum information processing~\cite{Walborn}. In particular, spatial quantum correlations between photon pairs generated with spontaneous parametric down conversion have been extensively studied over the last two decades~\cite{Teich,Monken,Fabre,Lantz1,Marcelo,Lantz2,Lugiato,shot1,shot2,shot3,Walborn,Perina,Perina1,EPR}. Owing to the conservation of momentum, or phase-matching, for the ideal case of a monochromatic planar pump beam and an infinitely thin non-linear medium, there are point-to-point spatial quantum correlations between the generated photon pairs in the far field~\cite{Monken}. However, for realistic experimental conditions, such as a Gaussian pump beam and a non-linear medium of finite length, the spatial correlations are spread out over a region in the far-field. The minimum size of this region is called the coherence area~\cite{Lugiato}. The size of the coherence area plays an important role in quantum imaging as it limits the resolution of the images~\cite{Kolobov,Brida}. Moreover, the presence of many coherence areas in the generated photons is a signature of the multi-spatial mode nature of the generated fields~\cite{Marcelo,Lantz2,Lett5,MarinoCoh,Ben}. Such a multi-spatial mode nature plays an important role in  quantum information, as it allows for parallel quantum information processing with each spatial mode playing the role of an independent quantum channel~\cite{RMP}.

In this paper, we compare two different techniques to characterize the size of the coherence area of bright entangled twin beams. The first technique is based on imaging entangled beams of light with a high quantum efficiency charge-coupled-device (CCD) camera and provides a direct measure of the size of the coherence area. The second technique employs a single photodiode for time-domain noise measurements of different spatial regions of the entangled twin beams selected with apertures. While the second technique provides an indirect measure, it is significantly easier to implement and does not require expensive equipment such as a high quantum efficiency CCD.  To date, this indirect technique has been used to verify the multi-mode nature of the fields ~\cite{Fabre,Marcelo,Lett5} and to provide a measure of how the coherence area changes as a function of the size of the pump beam~\cite{MarinoCoh}. Here, for the first time to the best of our knowledge,  we show that the time-domain technique can provide an accurate absolute measure of the size of the coherence area consistent with the direct technique.

\section{Experiment}

To perform the comparison between the two techniques, we first capture images of the entangled twin beams, which we call probe and conjugate, with an electron-multiplying charge-coupled-device (EMCCD) camera.  These images are then used to extract the size of the coherence area through cross~(auto) correlation measurements of the spatial fluctuations in the photocounts of the images of the probe and conjugate beams. We then replace the EMCCD camera with  slits of variable size and photodiodes, and analyze the noise of each beam in the time domain as a function of the transmission through the slits with a spectrum analyzer. We show that both techniques give the same estimate of the size of the coherence area, which validates the time-domain technique as well as the corresponding model used to extract the size of the coherence area~\cite{MarinoCoh}. Moreover, the time-domain technique does not suffer from many of the experimental complications associated with the spatial-domain technique.

To perform the experiments, we use bright twin beams generated with a four-wave mixing (FWM) process in a double-$\Lambda$ configuration in hot rubidium atomic vapor~\cite{Lukin, Lett1}. This process is based on a third-order nonlinearity and produces narrowband bright quantum correlated beams~\cite{Lett6,Howell} without the need of a cavity. This makes it possible to generate bright twin beams with a large number of spatial modes. As compared to entangled photon pair experiment, with these bright twin beams we are able to keep the level of quantum correlations fixed while independently controlling the number of photons \cite{Lett1,Glorieux,Kumar}.  This makes it possible to perform measurements with images acquired in a single shot. These properties provide significant advantages for quantum metrology, quantum imaging, and quantum information processing~\cite{Lett2,Lett3,Lett4,Pooser,Holtfrerich,Dowran}.

\begin{figure}[hbt]
\centering
\includegraphics[width=\linewidth]{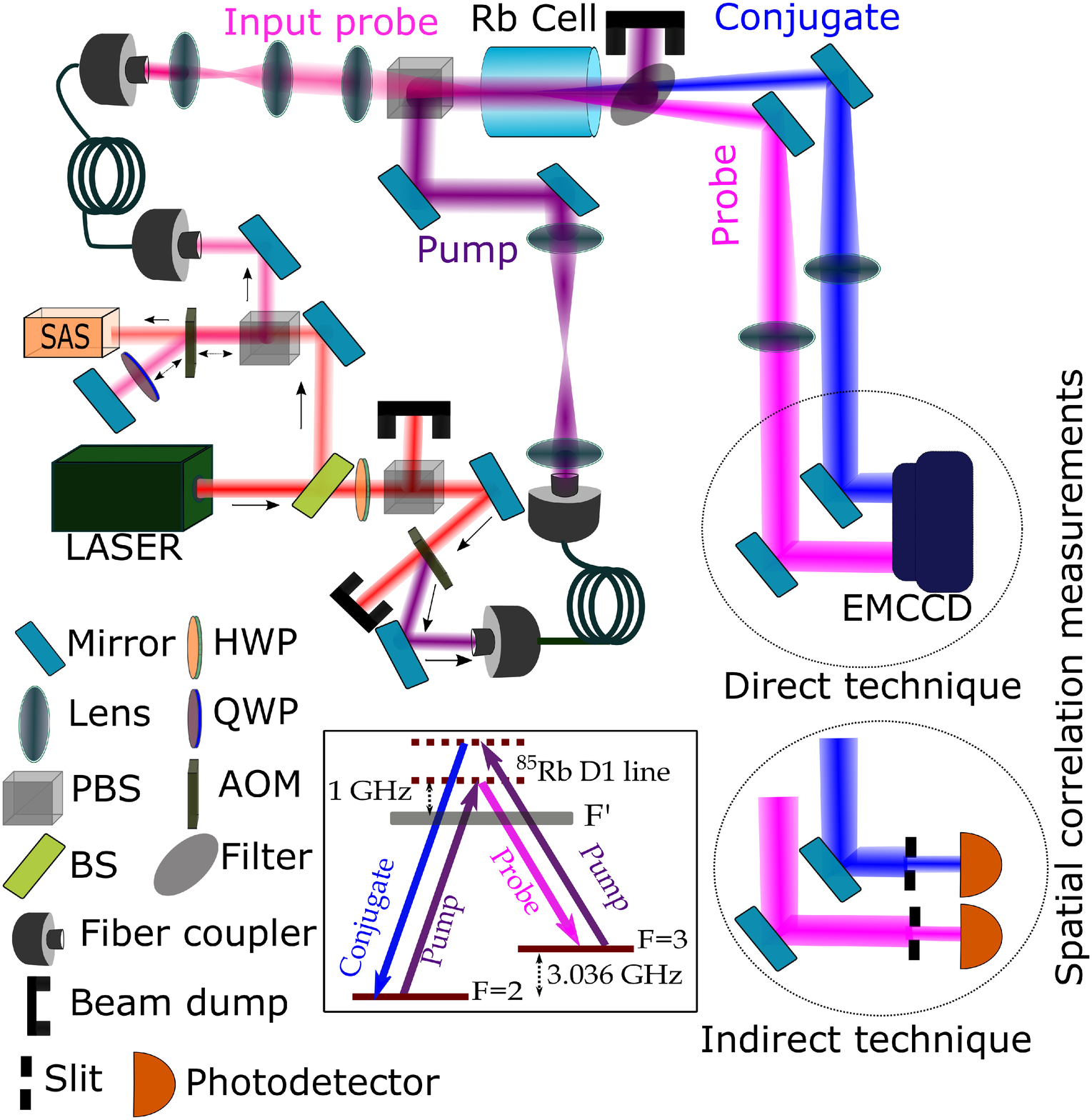}
\caption{Experimental setup. The size of the coherence area is characterized through two techniques.  In
the direct technique an EMCCD is used to acquire images and perform spatial correlation measurements. In the indirect technique the EMCCD is replaced by photodiodes and slits of variable size that are used to select different spatial regions of the twin beams. SAS: Saturated Absorption Spectroscopy, PBS: Polarizing Beam Splitter, BS: Beam Splitter, HWP: Half Wave Plate, QWP: Quarter Wave Plate, AOM: Acousto optic Modulator. The inset shows the double-$\Lambda$ energy level configuration in the $D1$ line of $^{85}$Rb used for the FWM.}
\label{fig:Fig1}
\end{figure}

A schematic of the experimental setup is shown in Fig.~1. A strong pump beam (power of 2.0~W, $1/e^2$ waist diameter of 4.4~mm) and an orthogonally polarized weak probe beam ($1/e^2$ waist diameter of 0.4~mm), both derived from the same Ti-Sapphire laser, interact at an angle of 0.4 degree with $^{85}$Rb atoms inside a glass cell heated to 110$^o$C. The inset in Fig.~1 shows a schematic of the double-$\Lambda$ energy level configuration on which the FWM process is based. In this process, the input probe beam is amplified and a new beam called the conjugate is generated, with the emission direction of these beams governed by the phase matching condition. The frequency of the laser is locked 1~GHz away from the atomic hyperfine transition $F=2$ to $F'=3$ of the $^{85}$Rb  D1 line through a saturated absorption spectroscopy setup. The frequency of the probe beam is down shifted by 3.04~GHz  with respect to the pump frequency with an acousto-optic modulator (AOM). After the cell, the pump beam is filtered with a polarization filter.

To study the spatial quantum correlations between the probe and conjugate beams in the far field, lenses are placed in the path of each of the beams after the cell to get the Fourier transform of the center of the cell onto the location of the EMCCD camera or the slits. However, due to a cross-Kerr effect between the pump and the probe and conjugate, the Fourier planes do not lie at the expected planes~\cite{Kumar}. To overcome this issue, we constructed an imaging system before the cell such that the Fourier transform of  an object in the input probe beam's path is formed at the center of the cell. This makes it possible to determine the optimum position of the lenses to get the actual Fourier planes at the detection planes after the cell by optimizing the image of the object at the location of the EMCCD camera or the slits.

\section{Results and discussion: Direct technique}

We first implement the direct technique by imaging the probe and conjugate beams with a high quantum efficiency EMCCD and subsequently computing the cross-correlation between the spatial intensity fluctuations of the two beams. While acquiring the images, the pump and the input probe beams are pulsed. The details of the pulse sequences and image acquisition with the EMCCD camera are described in Ref.~\cite{Kumar}. We record 200 images with multiple frames in each image. To obtain the spatial intensity fluctuations of the probe and the conjugate, we subtract two consecutive frames in each image. This allows us to eliminate the bright spatial profile of the twin beams that would otherwise dominate over the spatial fluctuations that contain the quantum correlations.

Due to phase-matching, the spatially correlated regions between the probe and conjugate beams are located diametrically opposite to each other with respect to the pump in the far field. Therefore, to calculate the cross-correlation, we rotate the image of the spatial intensity fluctuations of the conjugate beam by 180 degrees with respect to the corresponding image of the probe beam before the analysis. We then select a region of $80\times80$~pixels of the image of the spatial intensity fluctuations of the probe (conjugate) and scan it over a region of $120\times120$~pixels of the corresponding conjugate (probe) image to obtain the cross-correlation.  We then average the calculated cross-correlations over the 200 captured images to obtain the results shown in Figs.~2(a) and 2(b). The width of the spatial cross-correlation peak gives a measure of the coherence area of the twin beams. The obtained cross-correlation can be fitted with a two-dimensional Gaussian function, which allows us to obtain a full width at half maximum (FWHM) of both cross-correlation functions, shown in Figs.~2(a) and 2(b), of 12.4$\times$11.6~pixels along horizontal (X) and vertical (Y) directions, respectively.

\begin{figure}[htb]
\centering
\includegraphics[width=\linewidth]{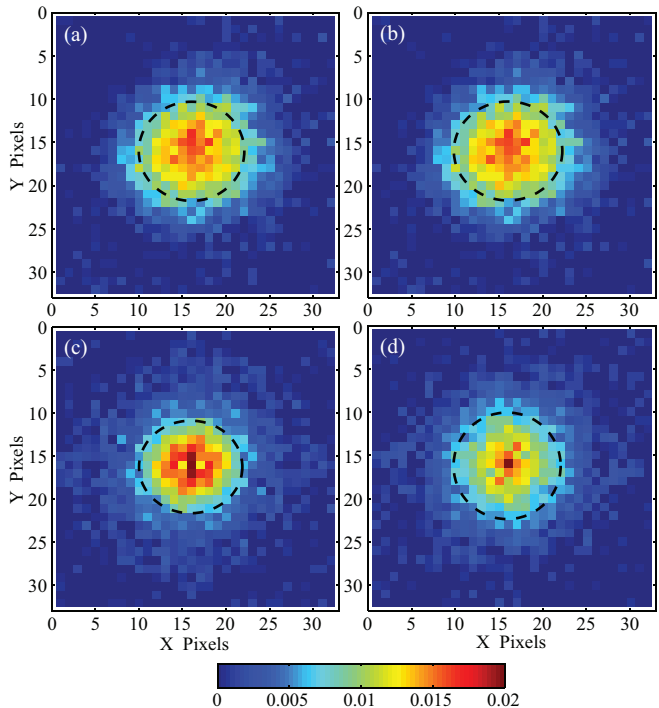}
\caption{Correlation functions. (a, b) Cross-correlation function between the intensity fluctuations of the probe and conjugate pulses when (a) the probe is scanned over the conjugate and (b) when the conjugate is scanned over the probe. (c, d) Auto-correlation functions of the intensity fluctuations of (c) the probe and (d) the conjugate. The dashed circles show the FWHM for the correlation functions.}
\label{fig:Fig2}
\end{figure}

The presence of a correlated region between the probe and the conjugate also plays a role on the spatial properties of the individual beams.  To show that this is the case, we calculate the  auto-correlation of the spatial intensity fluctuations of the probe and conjugate beams. In order to do this, we implement a similar analysis as the one used to calculate the spatial cross-correlations, except that now a region of $80\times80$~pixels of the  image of the spatial intensity fluctuations of the probe (conjugate) is scanned over the corresponding region of $120\times120$ pixels of the probe (conjugate) image. The auto-correlations that result from these calculations are shown in Figs.~2~(c) and 2(d). The FWHM of the auto-correlations for the X and Y directions are 11.4$\times$10.8~pixels, respectively, for the probe and 12.2$\times$12.4~pixels, respectively, for the conjugate. These values are consistent with the widths obtained from the cross-correlation measurements shown in Figs.~2(a) and 2(b) and show that the coherence area information can also be extracted from measurements of a single beam.

\section{Results and discussion: Indirect technique}

Given that the information of the coherence area is contained in each of the twin beams, it is possible to estimate its size through a noise analysis of only one of the beams. Here we present a time-domain noise analysis of the intensity fluctuations of different spatial regions of the probe or conjugate beams detected with a photodiode, we refer to this method as the indirect technique. We use a slit of variable size at the Fourier transform plane for either the probe or the conjugate and analyze the noise in the time domain for different transmissions through the slit. The noise in each beam can be quantified with the Mandel Q-parameter, defined as
\begin{equation}
Q=\frac{\langle(\Delta \hat{N})^2\rangle}{\langle \hat{N} \rangle}-1,
\label{Q}
\end{equation}
where $\hat{N}$ is the photon number operator. The Q-parameter defines the intensity noise normalized to the noise of a coherent state of the same intensity minus 1. For a coherent state $Q=0$, which corresponds to the shot noise limit.

The Q-parameter does not directly measure the coherence area; however, it can be used to determine if a given optical field is composed of a single spatial mode or of multiple spatial modes. For a single spatial mode, the Q-parameter varies linearly with transmission independent on whether the intensity of the beam is attenuated uniformly with a neutral density filter or by  cutting different spatial regions. On the other hand, for a multi-spatial mode beam, the behavior is not linear~\cite{Marcelo,Fabre}. As the number of spatial modes in a given field depends on the size of the coherence area, it is possible to obtain an estimate of the size of coherence area from the non-linear behavior of the Q-parameter as the beam is clipped.

We use the theoretical model that we developed in Ref.~\cite{MarinoCoh} to obtain an estimate of the absolute size of coherence area from the measurements of the Mandel Q-parameter. To date, this model has only been used to study relative changes in the size of coherence area and its accuracy in terms of the absolute size of the coherence area has not been validated. The current work shows, for the first time, that the size of the coherence area obtained for the indirect technique in combination with this model gives an accurate absolute measure consistent with the direct technique.

For the theoretical model the field operators are expanded in terms of a complete set of spatial basis modes. The functional dependence of the losses as a function of the spatial clipping is calculated for the basis modes, and this information is then used to calculate the Q-parameter. We assume that each of the basis modes used for the expansion can be treated as a single spatial mode in terms of the behavior of its Q-parameter with losses. Finally, the Q-parameter is normalized to its value at a transmission of one to obtain the normalized Mandel Q-parameter, $Q_N$,
\begin{equation}
Q_N= \frac{\sum_i \eta_i^2\langle\hat{n}_{i}\rangle}{\sum_i \eta_i\langle\hat{n}_{i}\rangle},
\label{QN}
\end{equation}
where $\eta_i$ is the transmission of  basis mode $i$ when clipping the beam and $\hat{n}_{i}$ is the number operator for mode $i$. In deriving Eq.~(\ref{QN}), we assume that all the basis modes have the same noise properties. This is a valid approximation as long as the beam does not occupy a significant portion of the spatial bandwidth. To extract the size of the coherence area from $Q_N$ as a function of transmission, we use a set of spatially localized modes that form a complete orthonormal basis set to perform the expansion. In particular, we use a set of 2-dimensional rect functions, which are equivalent to pixels in a CCD. Each of these modes are taken to be of size $2a\times2a$, where $a$ corresponds to the linear extent or radius of the coherence area.

In our experiment, to obtain the normalized Q-parameter for the probe (conjugate), we place a slit with variable size at the Fourier plane of the probe (conjugate) beam (shown as the indirect technique in Fig.~1). The field transmitted through the slit for different slit sizes is detected with a photodiode. To avoid diffraction effects from the slit, we use an imaging system that images the slit onto the photodiode. We divide the output photo-current from the photodiode into a low frequency (DC) and a high frequency (RF) component with a bias-tee, which has a cutoff frequency of 100~kHz.  We then analyze the noise power of the RF component with a spectrum analyzer at a frequency of 800~kHz, resolution bandwidth of 30~kHz, and video bandwidth of 100~Hz. For each transmission, we take 20 traces of the noise power with the spectrum analyzer. These traces are then used to calculate the mean noise power at each transmission. For the measurement of the transmitted power through the slit, we measure the optical power right before and after the slit with a power meter.

As shown in Eq.~(\ref{Q}), to calculate the Q-parameter, we need to normalize the noise power to the corresponding shot noise for each transmission. To measure the shot noise, we use a balanced detection scheme and record the difference noise for a coherent state as a function of incident power with a spectrum analyzer. The measured difference noise shows a linear relation to the incident power, characteristic of the shot noise. This provides a calibration of the shot noise for our detection system as a function of the incident power. With this calibration we obtain the required shot noise for the different transmitted powers through the slit. We finally use the measured noise power for the probe (conjugate) after clipping with the slit and the calculated shot noise to obtain the Q-parameter using Eq.~(\ref{Q}) for the probe (conjugate) beam when cutting the beams along the horizontal or vertical direction.

\begin{figure}[htb]
\centering
\includegraphics[width=\linewidth]{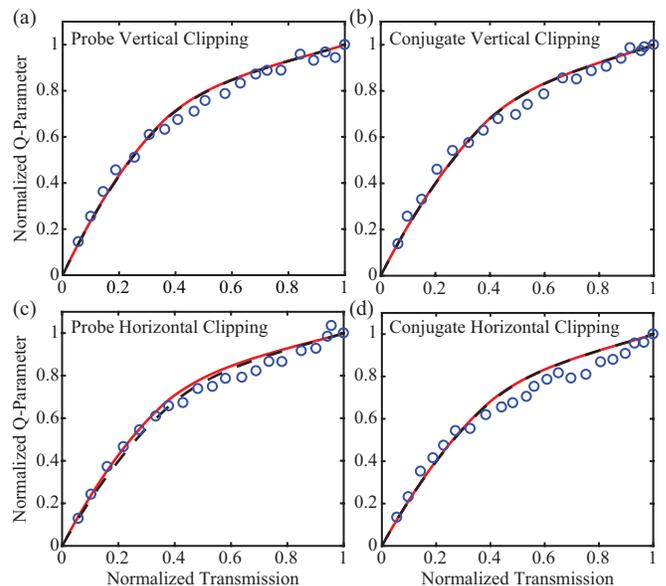}
\caption{Normalized Mandel Q-parameter for the probe (a,~c) and conjugate (b,~d) as a function of transmission through the variable size slit. The open circles represent the experimental data while the dashed black traces correspond to the best fit to the experimental data from our theoretical model. The solid red traces are the curves obtained from our theoretical model for the size of the coherence area measured with the auto-correlation for the direct technique.}
\label{fig:Fig3}
\end{figure}

Figure 3 shows the normalized Q-parameter obtained for the measured data and from the theoretical model as a function of transmissions for the probe, Figs.~3(a) and 3(c), and conjugate, Figs.~3(b) and 3(d). For each beam, the normalized Q-parameter when clipping in the vertical, Figs.~3(a) and (b), as well as in the horizontal, Figs.~3(c) and (d), directions are shown.  It can be seen that all the traces deviate from a linear behavior, which indicates the multi-spatial-mode nature of the beams. For each experimental set of data, we use the theoretical model to obtain the best fit to the data for the normalized Q-parameter based on a least square fitting technique of the normalized error. The fit for each case is given by the dashed black traces in Fig.~3. Our model has only one free parameter $a/\omega$, which corresponds to the ratio of coherence area radius ($a$) to the beam radius ($\omega$). From the best fit curves in Figs.~3(a)-3(d), we obtain the ratio $a/\omega$ as 0.26 (probe vertical), 0.29 (conjugate vertical), 0.29 (probe horizontal), and 0.29 (conjugate horizontal), respectively. The $1/e^2$ beam radii along the horizontal and vertical directions for our experiment at the position of the slit are $0.58\times0.57$~mm, respectively, for the probe and $0.57\times0.58$~mm, respectively, for the conjugate. This translates to a FWHM of the coherence area along the horizontal and vertical directions of $198\times174.5$~$\mu$m, respectively, for the probe and  $194.6\times198$~$\mu$m, respectively, for the conjugate. To obtain a direct comparison with the direct technique, we take into account the size of the pixels in the EMCCD ($16\times16$~$\mu$m)
to convert the estimated size of the coherence area to FWHM in pixel units. This results in a FWHM of the coherence area for the probe and conjugate of 12.4$\times$10.9~pixels and 12.2$\times$12.4~pixels, respectively. These results are in  agreement with the results obtained from the correlation measurements shown in Fig.~2.

Moreover, the solid red traces in Fig.~3 show the normalized Q-parameter obtained from the theoretical model for the size of the coherence area obtained from the auto-correlation measurements shown in Figs.~2(c) and 2(d). As can be seen from Fig.~3, the normalized Q-parameter for the fit obtain from the data with the indirect technique and for the one obtained from the measured size of the coherence area with the direct technique are almost identical.  This shows that it is possible to obtain an accurate absolute measure of the size of the coherence area with the indirect technique by using the theoretical model to fit the normalized Q-parameter.  Given that the model only has one free parameter, this provides a unique measure of the size of the coherence area.

\section{Conclusion}

In conclusion, we have compared two different techniques to estimate the size of the coherence area. In particular, we consider a ``direct'' technique based on a measure of the cross-correlation of the spatial intensity fluctuations of the twin beams and an ``indirect'' technique based on a noise analysis of different spatial regions of a single beam in the time domain. Results from both approaches give the same size of the coherence area and show that the absolute size of the coherence area can be estimated from a time-domain noise analysis of the twin beams. 

Our study shows that the significantly simpler indirect technique, which makes use of slits and photodiodes as opposed to an EMCCD camera for the direct technique, provides an accurate measure of the absolute size of the coherence area. In addition, the indirect technique offers significant advantages, as the EMCCD required for the direct technique is quite sensitive to scattered pump noise and imperfections in the optical system, which do not represent a significant issue for measurements done with the photodiodes. Moreover, for the direct technique it is necessary to cancel the gaussian spatial profile of the bright twin beams. This requires images of the twin beams to be acquired in rapid succession and thus a high quantum efficiency camera with a fast acquisition rate. As a result, the input pump and probe beams need to be pulsed and synchronized with the EMCCD. Overall, the indirect technique is immune to many of these complications given that the noise analysis is done in the time domain with a spectrum analyzer, which makes it possible to filter out technical noise at frequencies different than the analysis frequency.  The present study validates our theoretical model for estimating the size of the coherence area and shows that, in combination with the indirect technique, it provides an adequate characterization of the scale of the spatial correlations in the far field.\\
\\
This work was supported by the W.~M. Keck Foundation.

\end{document}